\documentclass{optica-article}

\journal{opticajournal} 

\articletype{Research Article}

\usepackage{lineno}

\begin{document}

\title{AI-Enabled sensor fusion of time of flight imaging and mmwave for concealed metal detection}

\author{Chaitanya Kaul,\authormark{1,*} Kevin J. Mitchell,\authormark{2,*,§} Khaled Kassem,\authormark{2,*} Athanasios Tragakis,\authormark{2} Valentin Kapitany,\authormark{2} Ilya Starshynov,\authormark{2} Federica Villa,\authormark{3} Roderick Murray-Smith,\authormark{1} and Daniele Faccio\authormark{2}}

\address{\authormark{1}School of Computing Science, University of Glasgow, Glasgow G12 8QQ, UK\\
\authormark{2}School of Physics \& Astronomy, University of Glasgow, Glasgow G12 8QQ, UK\\
\authormark{3}Dipartimento di Elettronica, Informazione e Bioingegneria, Politecnico di MilanoVia G. Ponzio 34/5, Milano 20133, Italy}

{\email{\authormark{*}These authors contributed equally to this work.}}
\email{\authormark{§}Kevin.Mitchell@glasgow.ac.uk}


\begin{abstract*} 
In the field of detection and ranging, multiple complementary sensing modalities may be used to enrich the information obtained from a dynamic scene. One application of this sensor fusion is in public security and surveillance, whose efficacy and privacy protection measures must be continually evaluated. We present a novel deployment of sensor fusion for the discrete detection of concealed metal objects on persons whilst preserving their privacy. This is achieved by coupling off-the-shelf mmWave radar and depth camera technology with a novel neural network architecture that processes the radar signals using convolutional Long Short-term Memory (LSTM) blocks and the depth signal, using convolutional operations. The combined latent features are then magnified using a deep feature magnification to learn cross-modality dependencies in the data. We further propose a decoder, based on the feature extraction and embedding block, to learn an efficient upsampling of the latent space to learn the location of the concealed object in the spatial domain through radar feature guidance. We demonstrate the detection of presence and inference of 3D location of concealed metal objects with an accuracy of up to 95\%, using a technique that is robust to multiple persons. This work provides a demonstration of the potential for cost effective and portable sensor fusion, with strong opportunities for further development.
\end{abstract*}

\section{Introduction}
Perception of a dynamic environment through a single (uni-modal) sensor inherently suffers from various limitations and vulnerabilities. For instance, consider a simple task of detecting the presence of objects in a room: a single RGB camera would be sufficient in many cases. However, such a system would be ineffective in low lighting conditions or in scenarios where the target objects are occluded. For the task of identifying the presence of a concealed metallic object on an individual, a mmWave radar transceiver is a well established choice. Objects illuminated by such Radio Frequencies (RF) exhibit varying reflectivities based on their material composition  \cite{Appleby2017,Wu2020,Taylor2015}. In this respect, mmWave radars can sense objects in a scene that are occluded, either due to the occlusion being transparent to these RF signals, or by the multipath nature of RF signals reflecting from diffuse surfaces \cite{Sen2011,Turpin2020,Turpin2021,Vakalis2019,Abbasi2023}. Commercial RF transceivers however, lack the ability to perform conventional imaging of a scene which may be an important criteria in real-world use cases. A combination of multiple sensors with specific imaging characteristics provides complementary information about a scene, which makes multi-modal data acquisition setups highly desirable for many applications. 

We propose a taxonomy of existing multi-modal information processing systems and argue that within the context presented, they can be broadly categorized into two groups - structured and hybrid sensing. In the former, data acquisition systems are generally comprised of various camera setups, such as RGB, depth and Light Detection and Ranging (LiDAR), which all spatially resolve the scene with a predetermined resolution. Such setups provide information about the structural nature of the scene in 2D and 3D. Hybrid sensing acquires complementary sets of data, where the sensors can be combinations of cameras, Lidars, Wi-Fi, radar etc. For example, a depth camera and radar set up can acquire different properties about the scene, and the data from one sensor complements the information provided by the other as they both capture different characteristics of what is being sensed without obvious overlap.

Technological innovation and commercialisation of the security sector opens the door to potentially invasive mass-monitoring on a global scale. In particular, screening the general public for illicit weapons and devices introduces several complications which are exacerbated by human error and risk. Most modern metal detection schemes rely on electromagnetic induction, mmWave reflection, or X-ray imaging. Metal detection scanning with an electromagnetic wand and performing pat-downs requires time, risks the introduction of discrimination and close proximity monitoring is an inherent safety risk. The presence of walk-through metal detection scanners has become commonplace in airports, transport hubs, stadia and other public thoroughfares, their use growing among schools and other private sites. Whilst their performance ranks very high, they are not infallible, and together with their great cost and a tendency for such scanners to bottleneck people traffic in major thoroughfares, the result is a security solution which does not satisfy all use-cases appropriately. The mmWave technology employed in airport security scanners has also long been under scrutiny for its ability to image through clothing, prompting privacy concerns \cite{Williams2005,ACCARDO2014198,rosenberg1998passive,Martin2007}. This suggests that there is  need for a technology that can remotely screen for metallic objects, whilst still allowing people to maintain their freedom and privacy in public without security bottlenecks and conventional image-capturing cameras.

In between these frequencies of operation sit mmWave imaging systems. Cheaper than X-rays, non-damaging to biological tissues, and offering quasi-optical spatial resolution, mmWave imaging is also preferential for imaging humans over prolonged doses. 
At the core of this technology lies the principle of leveraging the distinct reflectivities of different materials when subjected to mmWave radiation. This modality allows for differentiating between materials, such as skin and metal, based on their varying reflection properties \cite{hallbjorner2013improvement}. Additionally, mmWave signals can easily penetrate clothing \cite{yamada2005radar}. 

The state of the art mmWave-based solutions for metal detection rely on scanning gates. Several approaches have been developed that integrate mmWave scanning with AI for improved detection fidelity. Evolv Technology has integrated AI with bespoke mmWave/optical active imaging to produce scanning gates that can process up to 900 people per hour, exceeding CT-scan technologies \cite{baker2019can}. IPVM claims that their AI can enhance existing gates and report similar detection accuracy to Evolv's bespoke system \cite{ipvm}. Sequestim Ltd uses AI to enable passive walk-through scanning, i.e. seeing metal objects by the shadow they cast on the natural terahertz/infrared emission of the human body \cite{sequestim}. Despite the promise of AI in concealed metal object detection, there is still room for improvement. Reported incidents of undetected knives brought into schools through AI-enabled scanning gates has raised questions over their efficacy \cite{BBCNews,Vizitiu2024,Hankin2011,Bhatt2018,nirgudkar2024beyond}.

Recent works have shown that it is possible to accurately detect the presence of concealed metal on a person in a scene using only an mmWave radar transceiver \cite{mitchell2023mmsense,mmsensecite1,mmsensecite2,mmsensecite3}. A common limitation is the requirement to scan  a person, which may not capture the intricate details of the structure and spatial properties of the scene. 

In this work, we avoid the pitfalls of uni-modal radar sensing by adding a structural guidance to the system through a depth sensor. We propose and demonstrate that adding this additional structural guidance in the feature space is essential to add spatial context to the task of AI-assisted metal object detection. We collect multiple multi-modal datasets with multiple radars and depth cameras, to show that our technique generalizes across various permutations of radar and cameras and is not dependent on any particular combination of sensors. Finally, we propose a late-feature fusion model that first extracts features and embeds them into a high dimensional latent space to create representations of the multi-modal data. Following this, the model learns to optimize a combination of them together using back-propagation to detect the presence of a metal object. Our experiments demonstrate the ability of multi-modal sensor fusion to be used as an effective  detection system compared to single-radar setups.
\section{Methods}

\begin{figure}[htb]
\centering
\includegraphics[width=0.8\linewidth]{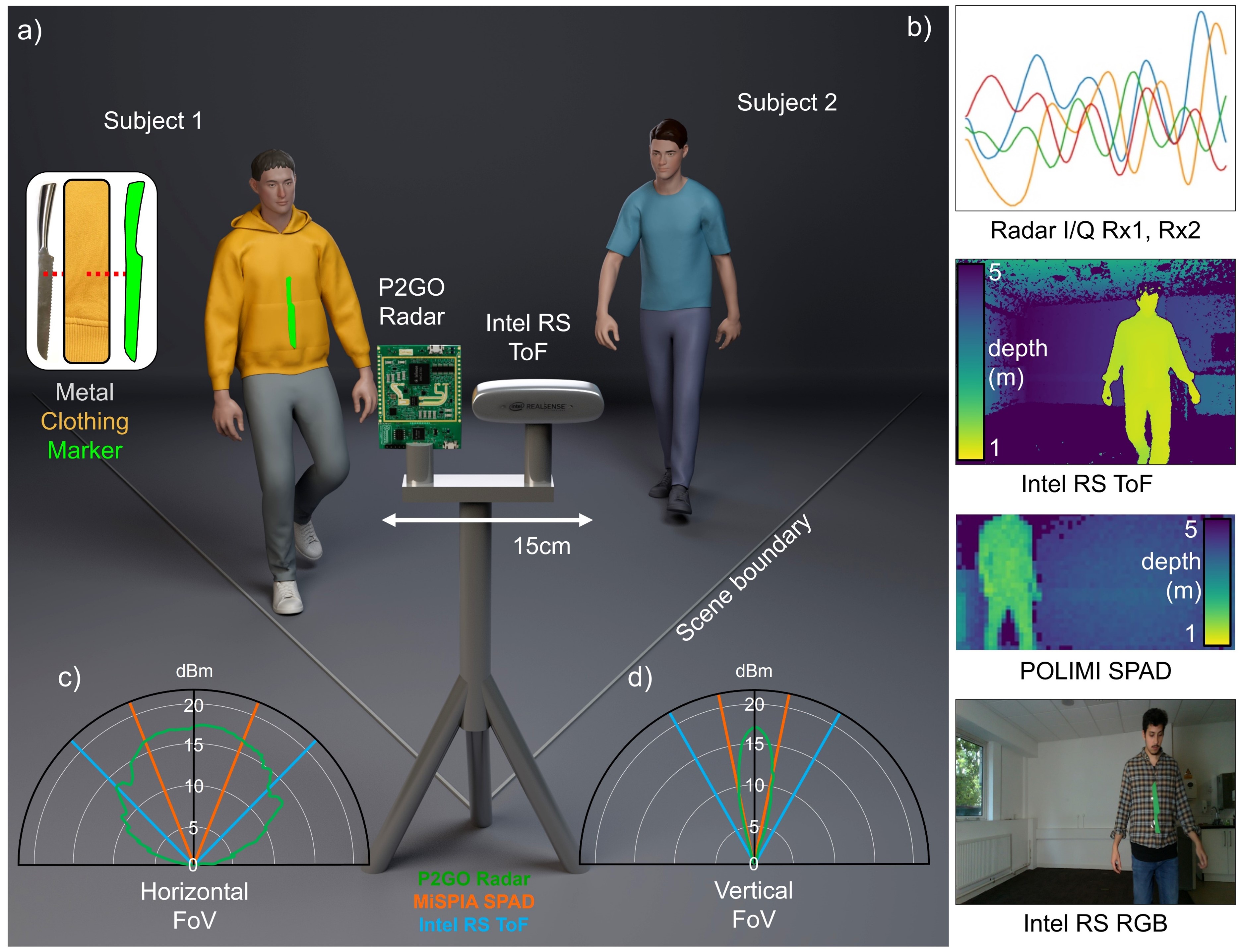}
\caption{Experimental setup for radar and depth camera-based concealed metal object detection. The setup in \textbf{a)} shows one or two subjects walking in view of the devices, with one subject concealing a knife beneath their first layer of clothing. The knife location is labelled with a green paper marker for training using the RGB camera data. The acquisition data modalities are listed in \textbf{b)}. Specifically, the intermediate frequency Radar signal, depth (either Intel RealSense or POLIMI SPAD camera) and RGB color images. The relative field of views (FoVs) for each device are shown in \textbf{c), d)} in both horizontal and vertical axes. Intel RS RGB image and all depth maps throughout depict the authors, with permission. 3D render courtesy of Diana Kruhlyk.}
\label{fig:experiment_procedure}
\end{figure}

Consider the setup shown in figure \ref{fig:experiment_procedure}: the radar and 3D depth camera are paired together before a static scene (approximately $5\times5$ metres). The former is an Infineon Xensiv Demo Position2GO 24GHz radar transceiver, with 1 transmitting and 2 receiving antennas capable of tracking multiple targets up to $\sim$20\:m using fast chirp FMCW radar technology \cite{FMCWRao,Skolnik1981,P2GOManual}. The latter is an Intel Realsense D435 stereo depth camera, which uses a pair of ultrawide sensors 50\:mm apart to calculate depth from stereo images. Furthermore, the Intel D435 includes an integrated colour (RGB) sensor which is co-registered to the depth data as a reference for training only \cite{Corporation2018}. 

To investigate the potential of SPAD (Single Photon Avalanche Diode) array technology, we also replaced the Intel depth camera with an indirect Time-of-Flight (iTOF) SPAD camera, developed at the Politecnico di Milano (POLIMI) \cite{Bellisai2013,Zappa2013,Bronzi2014}.
This SPAD camera generates a $64\times32$ resolution depth map of a scene with a  $20^{\circ}\times40^{\circ}$ field of view. This more closely matches the radar used. This SPAD-based solution opens the door for ultra low-light depth sensing, and different modes of operation could also provide access to the time dimension for single photon applications.

\subsection{Experimental Procedure} 

During data acquisition, subjects removed all metals from their person, and concealed a 20 cm long steel knife underneath their first layer of clothing, specifically on the chest. The location of the knife was labelled by affixing a green paper marker on top of the clothing, to be seen only in the RGB reference image whilst not affecting the other modalities. The procedure used for data collection and deployment of the metal detecting system was borne out of our previous work in generating 3D spatial images using temporal data \cite{Turpin2020,Turpin2021} in which a radar transceiver and depth camera pairing was used to gather data of a human subject moving through a static scene. 

The experiments were conducted with the sensor rig that was activated to stream simultaneously, whilst the subjects moved through the space at walking pace for between 2-6 sets of 3000 frames (limited by RAM storage limitations) at $\sim$20Hz. These data sets included various configurations e.g. subject with/without object concealed or multiple subjects. For simplicity, the subject always faced the sensor rig, whilst their randomised movements were intended to encompass all likely positions and velocities. All three data streams were then saved to file before post processing and training. During deployment of the trained system on test data, only the radar and ToF depth maps were used. 
We then input the depth image and raw radar I/Q signals to our dual input neural network to predict segmentation masks.

\subsection{Network Architecture}
\begin{figure}[htbp]
\centering
\includegraphics[width=1\linewidth]{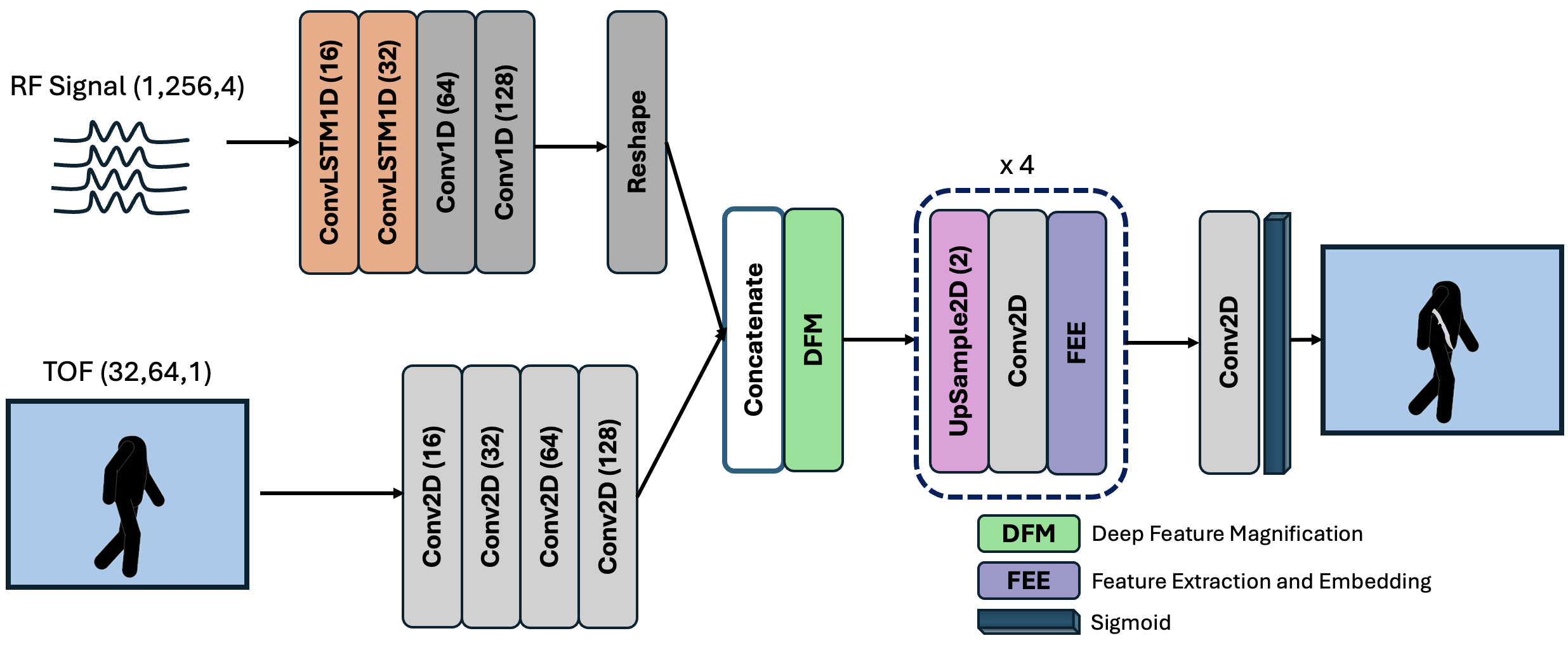}
\caption{Our neural network archtecture, $mmSense_{AF}$ for concealed metal detection. We process the radar data as sequences using ConvolutionalLSTM blocks and create embeddings from the spatial depth image using convolutional blocks with large receptive fields. After concatenating the embeddings from both modalities, we extract joint concepts from them using the deep feature magnification block. We then use a convolutional decoder coupled with a feature extraction and embedding module to upsample this encoding to generate the output mask.}
\label{fig:arch}
\end{figure}

Our neural network is presented in figure \ref{fig:arch}. Firstly, we create a dataset $\{\mathbf{X_i}, \mathbf{X_r}, \mathbf{Y}\}$, where $\mathbf{X_i}$ are the input Time-of-Flight (ToF) images, $\mathbf{X_r}$ is the input radar data and $\mathbf{Y}$ is the corresponding binary segmentation mask denoting the location of the knife in the ToF image. All SPAD ToF images are used in their native $32\times64$ resolution; the RealSense TOF images are downsampled from native to $48\times64$. Following this, we replace $0$-depth pixels with the minimum pixel value in the scene to create a smooth image without discontinuities which can cause gradient instabilities in training the neural network. After this we standardize the images by first subtracting the mean pixel value and then normalizing the image with the maximum value. The radar data is normalized by dividing the radar sequence by the maximum in the sequence to scale the intensity values to a $[0, 1]$ interval. 

\subsubsection{Implementation}

Our model is trained using TensorFlow 2.12.0 for a maximum of 100 epochs. We use a batch size of 64 for all our experiments. We use the Adam optimizer with an initial learning rate of 1E-3 which we then reduce on a plateau to a minimum value of $1e-6$. We run all our experiments on one Nvidia A5000 GPU. All weights for our model are initialized from a random normal distribution. We use the binary cross entropy loss to create a measure for the difference between the distributions of the predicted and ground truth labels. 

\subsubsection{mmSense$_{AF}$}

\begin{figure}[htbp]
\centering
\includegraphics[width=0.8\linewidth]{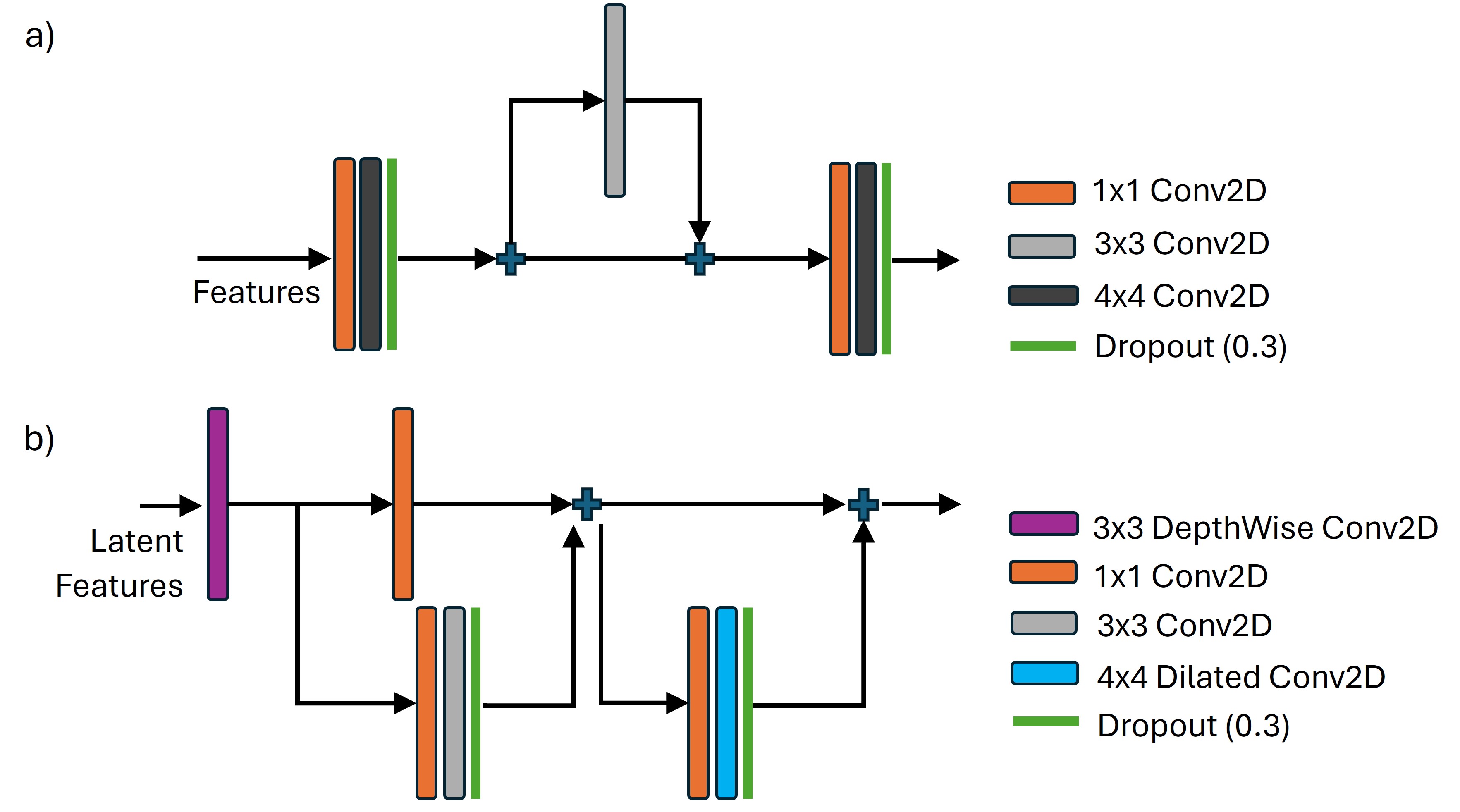}
\caption{a) The structure of the deep feature magnification block that takes the concatenated features from both modalities as an input and learns a relation between them by focusing on relevant features using increasing convolutional kernels and receptive field sizes in the convolutional block. b) The feature extraction and embedding block processes the upsampled latent features with increasing convolution kernel sizes to learn to correlate the location of the concealed object with the depth features by processing the encodings at varying receptive fields of the convolution kernel.}
\label{fig:dfm}
\end{figure}

Our neural network architecture, called $mmSense_{AF}$ (\textit{Auto-Focus}), is a dual-channel convolutional encoder-decoder structure that iteratively learns to focus on the combined latent features from the radar and TOF modalities in order to extract the 3D location of the radar mask in the TOF image. It consists of distinct encoders for both modalities and processes them based on the properties of the individual mode, extracting features from the radar signal and TOF images and outputting their latent representations. Following feature extraction, the two embeddings are concatenated across the depth axis to create a latent representation which serves as a joint embedding of the data. We process this joint embedding to create a true latent space of combined radar and TOF features using the Deep Feature Magnification block (DFM), as shown in figure \ref{fig:dfm}a). The outputs from DFM are then upsampled and passed through a convolution layer to generate upsampled features. We then process these features using a Feature Extraction and Embedding (FEE) block (figure \ref{fig:dfm}b)). The output from the last FEE block is passed into a convolutional layer with a \texttt{sigmoid} activation function to generate a per-pixel probability of the location of the concealed metal object in the TOF scene. 

\textbf{Radar Encoder.} The radar encoder consists of two \texttt{ConvolutionalLSTM1D} layers followed by three \texttt{Conv1D} layers. We pass a $1 \times 256 \times 4$ radar input to the the first \texttt{ConvolutionalLSTM1D} layer to extract time dependant features from the input data sequence. The second layer further processes this data and converts the sequences into convolution compatible features. The following two convolution layers then create high dimensional embeddings of the data using a stride of $2$ to simultaneously reduce the feature map size. All convolutional kernels in this layer have a large receptive field of $(1,7)$ in order to incorporate more time dependent information and a larger spatial context to compute the features for the next layers. The third convolution layer is used to learn the final embedding of the radar data. The output filter maps $16 \rightarrow 32 \rightarrow 64 \rightarrow 64 \rightarrow 128$. All layers use a \texttt{ReLU} activation function. The output from the last \texttt{Conv1D} layer is reshaped to resemble the output shape of the TOF features. 

\textbf{TOF Encoder.} The TOF encoder consists of 4 \texttt{Conv2D} layers with zero padding and the \texttt{ReLU} activation function. Each convolutional layer uses strides of varying receptive fields to create a reshaped latent representation of $4 \times 4 \times 128$ features making it compatible with the reshaped radar features. We use large receptive fields of size $(1,7)$ in this encoder as depth images tend to be smoother than RGB images and have less intensity transitions. Further, given that our goal is to learn global properties of the scene and not local features about the individual in the scene (which are not visible in a TOF image), a large receptive field allows us to capture the relation of the large objects in FOV of the camera. The output filter maps $32 \rightarrow 64 \rightarrow 64 \rightarrow 128$.

\textbf{Latent Fusion.} We coarsely combine the radar and TOF latent fatures via a simple concatenation operation. We then pass this joint embedding into the DFM block \cite{dfmfee}. The DFM block magnifies the relationships between the features of the two modalities. The concatenated features are first passed through a \texttt{DepthwiseConv2D} layer. This first learns individual features for all $256$ concatenated filter maps in the feature representation with a spatial receptive field of $3 \times 3$ followed by a $1 \times 1$ convolution across the depth of the features to aggregate information from both modalities into a learnt joint representation. These features are then incrementally processed by convolutions of varying receptive fields, $1 \times 1$, $3 \times 3$, $1 \times 1$ and $4 \times 4$ to create spatially consistent representations along with extracting any subtle feature dependencies that exist between the representations of both modalities. We also use a $4\times4$ convolution with a dilated receptive field to learn larger spatial relationships between the data which facilitates the interlinking of dynamic structures in the feature space across a larger region providing larger global context to the features. We use dropout layers with a rate of $0.3$ throughout to prevent overfitting. All convolution layers in the DFM block use \texttt{BatchNormalization} followed by the \texttt{ReLU} activation function. 

\textbf{Decoder.} We decode the latent fusion output by incrementally upsampling it by a $2 \times 2$ factor and processing the upsampled output with a convolution layer. These features are passed to the FEE block \cite{dfmfee} to extract further spatial correlations in the features at the different upsampled scales. The FEE block performs consecutive $1 \times 1$ and $4 \times 4$ convolutions to the features. Similar to the DFM block, having a small $1 \times 1$ receptive field builds local features across all filter maps while keeping the parameter size of the model low. Following this with a large $4 \times 4$ receptive field size aggregates the features along with a larger spatial context. As the location of the hidden object is only a few pixels in size, extracting small receptive field features across depth and aggregating them with a larger spatial context helps to localization by looking at the features at multiple receptive fields. Similar to DFM, we use dropout layers with a rate of $0.3$ throughout to prevent overfitting. All convolution layers in the FEE block use \texttt{BatchNormalization} followed by the \texttt{ReLU} activation function. The output from the last FEE block is processed by a convolution operation with a receptive filed of $1 \times 1$ and a \texttt{sigmoid} activation function to generate per pixel probabilities for the presence of a concealed object on the person in the scene. 

\section{Results}

\begin{table}[h]
\centering
 \begin{tabular}{c| c c c c} 
 \hline
Regime & Accuracy (\%) & Sensitivity (\%) & Specificity (\%) & Precision (\%) \\ 
 \hline
SPAD $\mathtt{1P}$ & 94.5 & 93.4 & 93.8 & 96.6 \\ 
SPAD $\mathtt{2P_1}$ & 85.6 & 91.0 & 61.6 & 92.2 \\ 
SPAD $\mathtt{2P_2}$ & 70.0 & 70.4 & 69.6 & 74.6 \\ 
TOF $\mathtt{1P}$ (Wide FOV) & 67.4 & 64.4 & 75.6 & 60.3\\ 
\hline
\end{tabular}
  \vspace{0.2cm}
 \caption{Our results table provides quantitative results on the different data acquisition regimes in terms of standard metrics. All results are generated using our latent space feature fusion model with the Infineon P2GO radar and a depth camera (SPAD refers to the POLIMI camera and TOF Wide FOV refers to the Intel RealSense D435).}
 \label{tab:mainresults}
\end{table}

\begin{table}[htbp]
\centering
 \begin{tabular}{c| c} 
 \hline
 SPAD 1P & Accuracy (\%) \\ 
 \hline
Depth Only & 50.4 \\ 
Radar Only & 74.6 \\ 
Radar+Depth & 94.5 \\
Radar+Depth (Wide FOV) & 67.4 \\ 
\hline
\end{tabular}
  \vspace{0.2cm}
 \caption{Demonstrating the need for sensor fusion. Our results demonstrate that fusing the spatial and radar modalities allows us to learn correlations between both domains and creating a more accurate regime for concealed metal detection.}
 \label{tab:ablation}
\end{table}

\begin{figure}[ht]
\centering
\includegraphics[width=0.8\linewidth]{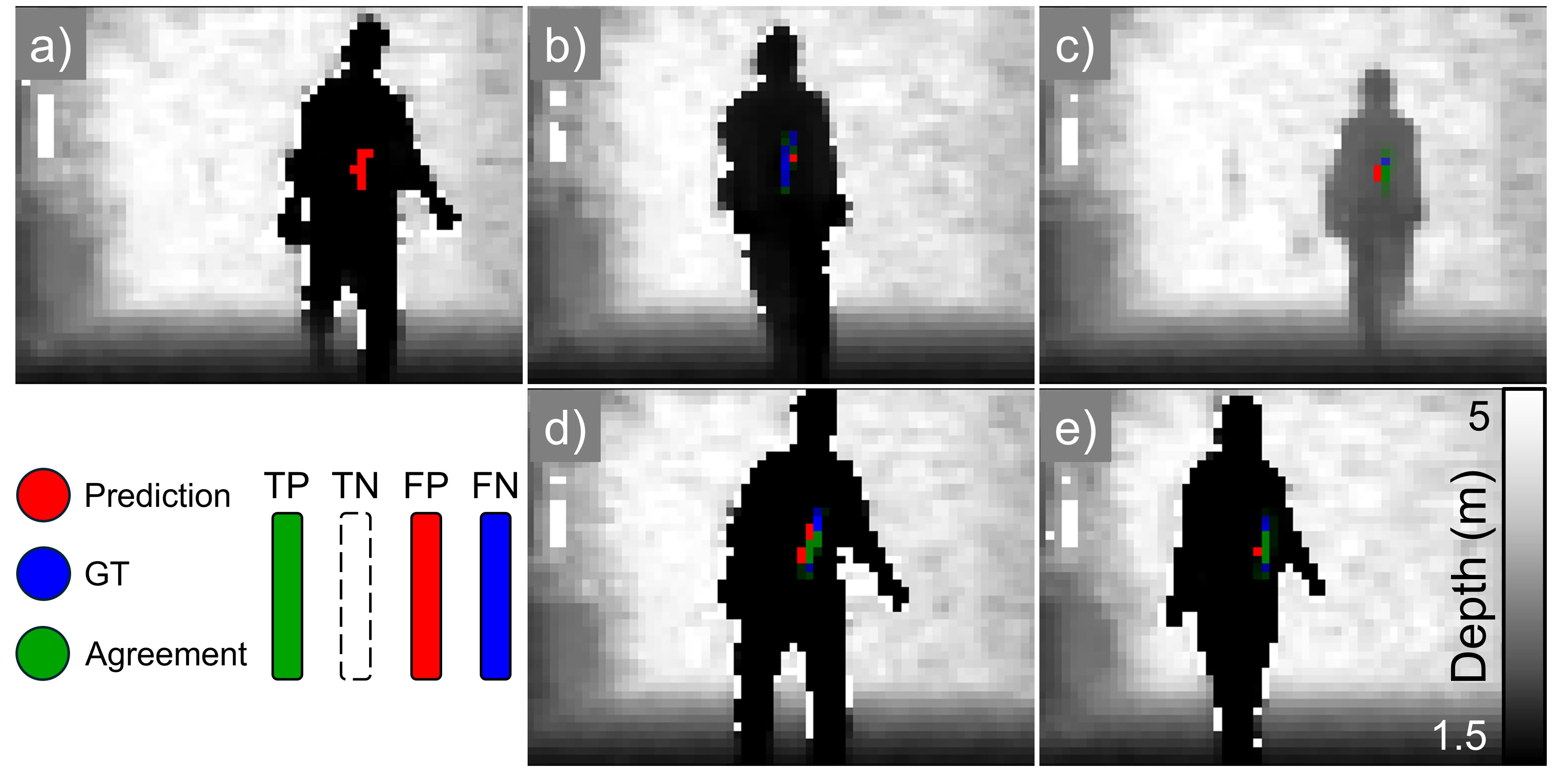}
\caption{TOF 1P (Wide FOV) visualizations. Prediction (red) and ground truth (blue) are overlayed to depict agreement (green) which represent the standard F-scores in each frame. a) High False Positive, b) High False Negative, c,e) High Agreement, d) Translated prediction. Shown images are samples of video frames from a test set which comprises a prediction mask overlayed on the original depth frame, for the P2GO radar and the Intel Realsense depth camera with one person in the scene. Visualization 1 in the supplementary materials.}
\label{fig:tof1p}
\end{figure}

\begin{figure}[ht]
\centering
\includegraphics[width=0.9\linewidth]{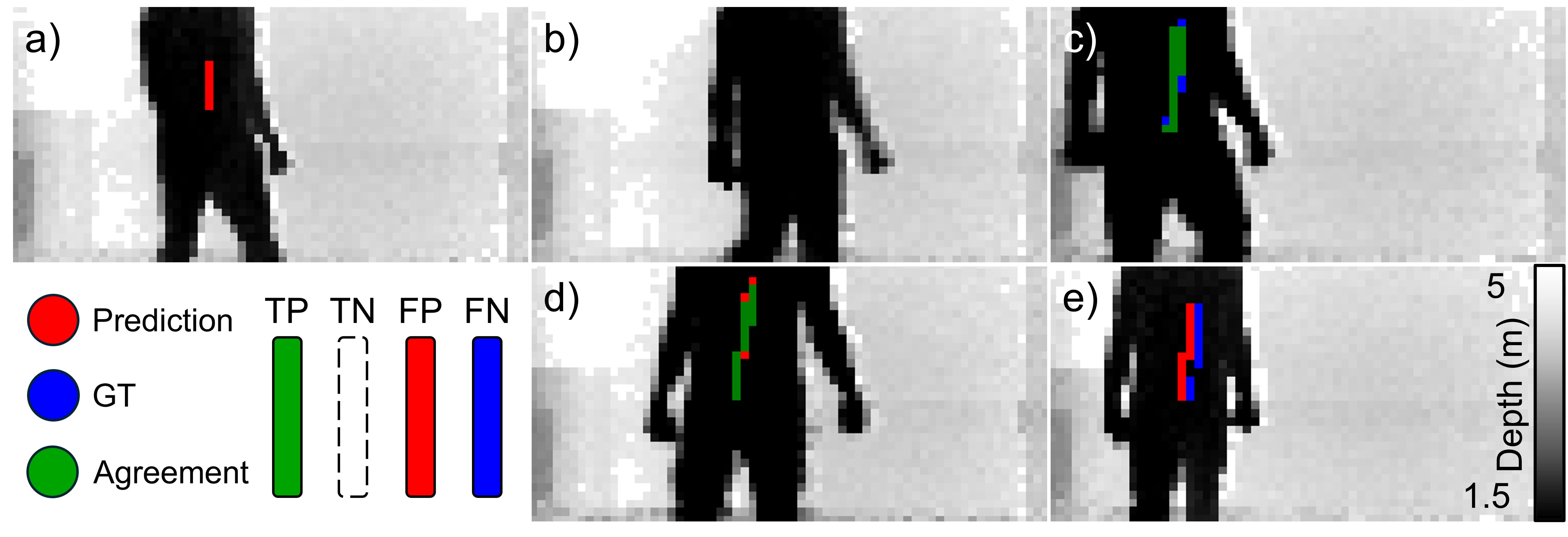}
\caption{SPAD $\mathtt{1P}$ visualizations. a) High False Positives, b) Negative Label, c,d) High Agreement, e) Translated prediction. Shown images are samples of video frames from a test set which comprises a prediction mask overlayed on the original depth frame, for the P2GO radar and the POLIMI SPAD camera with one person in the scene. Visualization 2 in the supplementary materials.}
\label{fig:spad1p}
\end{figure}

\begin{figure}[ht]
\centering
\includegraphics[width=0.9\linewidth]{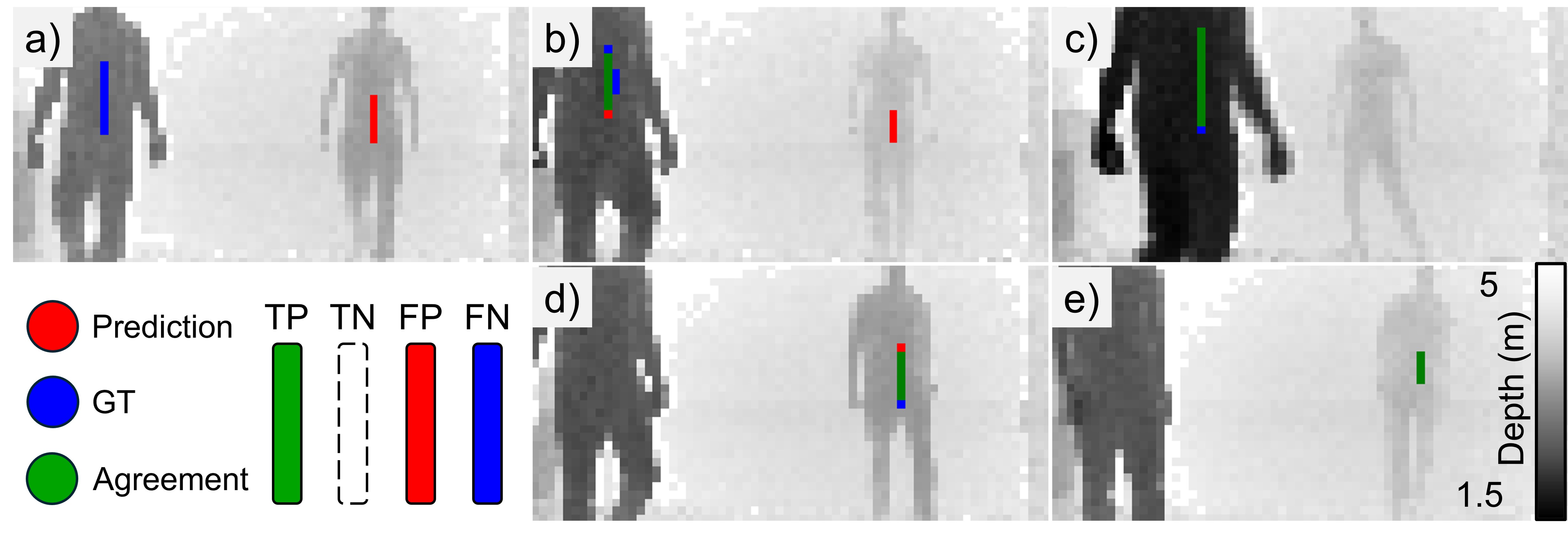}
\caption{SPAD $\mathtt{2P_2}$ visualizations. a) Mis-labelling, b) Semantic issue, c-e) High Agreement. Shown images are samples of video frames from a test set which comprises a prediction mask overlayed on the original depth frame, for the P2GO radar and the POLIMI SPAD camera with two people in the scene. Visualization 3 in the supplementary materials.}
\label{fig:spad2p1}
\end{figure}

We demonstrate the effectiveness of our system quantitatively in Table \ref{tab:mainresults} and qualitatively in figures~\ref{fig:tof1p}, \ref{fig:spad1p} and \ref{fig:spad2p1} (still frames from supplementary videos S1,S2 and S3 respectively, with the acquired datasets and architecture publicly available in \cite{dataset}). We also show the necessity of sensor fusion to integrate spatial information into the feature processing through our results in Table~\ref{tab:ablation}. We report standard metrics, namely, accuracy (agreement), sensitivity (True Positive (TP) Rate), specificity (True Negative (TN) Rate) and precision (Positive Prediction Value) of our model's ability to predict the presence of the concealed metal object (knife) in a 3D scene. We define agreement with the ground truth, GT (shown in green), when the centre weight position of Prediction is within 5 pixels the GT position. Here, row 1-3 present the values for the POLIMI SPAD camera and the P2GO radar: row 1 probes the concealment from one person, row 2 assumes a metal object is present and identifies which person has it; row 3 combines both concealment and identifying the person.  Row 4 shows the same test as row 1, but for the Intel RS depth camera instead, which has a larger field of view and extends to regions where the P2GO radar weakens in strength. We report strong detection capabilities for the single person case, and promising values in the multiple person case. 

Table \ref{tab:ablation} shows the value in fusing the two modalities together. We note that an accuracy of 50\% using just the depth modality understandably amounts to a random guess as the concealed metal is not visible in the spatial domain, whereas the Radar+Depth result nears 95\% compared to 74.6\% for the radar alone. This also demonstrates that the depth image contributes little towards the metal detection, but serves as a {discriminator} for the network to correlate spatial information with the radar data, to train on the radar data more effectively.

Figures \ref{fig:tof1p}, \ref{fig:spad1p} and \ref{fig:spad2p1} visualize the results in the one person and two person cases when pairing the Infineon P2GO and the POLIMI SPAD camera camera as our data acquisition set up. We show visualizations of high agreement, high true or false positive results as well as predictions where the presence of the metal is translated in the image, but the structure of the metal is predicted correctly. 

\section{Conclusions and Discussions}

In this work we have demonstrated the first promising results towards 3D remote detection of concealed metal objects using radar and depth data fusion. We proposed a feature fusion and processing framework based on a deep encoder-decoder neural network architecture capable of extracting relevant features at multiple receptive fields of the input to locate the presence of concealed metal in a 3D scene. We have demonstrated the effectiveness of such a system for static scenes containing single and multiple subjects. However, it is important to underline the considerable challenges faced when generalising this technique to a wider range of scenarios. 

Firstly, we chose to focus only on a static fixed scene within which to detect. This greatly simplifies the problem by limiting the variables to the subjects and the presence of metal objects on their person. In order to generalise to different background scenes, a much larger training dataset is needed. The number of participants in the scene was also constrained to one or two, as the transceiver used features 1Tx and 2Rx antennas and an equivalent depth resolution of 70cm -- more subjects in the scene would convolute the signal received and make identification more difficult. Alternative transceivers with more angle and depth resolution are available  which would allow for up to 5 subjects in the scene to be tracked through radar alone \cite{electronics12020308,Huang2021,Pegaroro2021,Xu_2022} . 

We also note the impact of specular reflections of the radar signal from the metal objects. We found that flat surfaced metals tend to reflect the signal without any discernible scattering. A consequence of this is the ability to detect flat metal objects at distances of multiple metres due to the higher signal to noise, but only when the surface normal of the flat metal surface and the radar optical axis deviate by less than $5$--$10^{\circ}$. To maintain simplicity and relevance for this proof of concept, we ensured that the subjects faced the detector throughout the acquisition (akin to use cases in a public thoroughfare), and selected a steel bread-knife as the metal object due to its multiple angled surfaces and its relevance to real-world concealed carry. These challenges may be mitigated by multiple separated detectors, or curved detection rings which provide multiple viewpoints to the subjects. 

This work seeks to highlight the untapped potential of portable commercial radar and depth sensing technology for metal detection. The approach provides a discreet, cost-effective and safe means to deploy metal detection in public and private spaces. We have observed that, provided there is sufficient training data and the availability of powerful AI models, the potential exists to generalise this concept to wearable technology which can run in real-time with background independence, which could be part of a step change in surveillance and security of the future.
\newline
\newline See visualizations 1-3 for supplementary videos.

\section*{Acknowledgements}
D.F. acknowledges funding from the Royal Academy of Engineering Chairs in Emerging Technologies programme and the UK Engineering and Physical Sciences Research Council (grant n. EP/T00097X/1).
R.M.S. and C.K. received funding from EPSRC projects Quantic EP/T00097X/1
and QUEST EP/T021020/1 and from the DIFAI ERC Advanced Grant proposal 101097708, funded by the UK Horizon guarantee scheme as  EPSRC project EP/Y029178/1. This work was in part supported by a research gift from  Google.

\section*{Author contributions statement}
D.F. conceived the experiments. K.J.M., K.K. formulated the experiments and conducted the data acquisition. C.K. K.K. A.T. and R.M-S. provided the neural network. C.K. K.K. and A.T. ran the experiments on the datasets and calculated the results. C.K. K.J.M and K.K. wrote the paper. F.V. developed the POLIMI iTOF SPAD Camera. All authors reviewed the manuscript.

\section*{Disclosures}
The authors declare no conflicts of interest.

\section*{Data Availability}
The data underlying the results presented in this paper is available in \cite{dataset}.

\bibliography{metal_detection_main_paper}

\end{document}